\documentclass[]{rsproca_new}%%%%where rsproca is the template name

\begin{document}

%%%% Article title to be placed here
\title{Multiple scales analysis of slow--fast quasilinear systems}

\author{%%%% Author details
G. Michel$^{1}$ and G. P. Chini$^{2}$}

%%%%%%%%% Insert author address here
\address{$^{1}$Laboratoire de Physique Statistique, \'Ecole Normale Sup\'erieure, CNRS, Universit\'e P. et M.
Curie, Universit\'e Paris Diderot, Paris 75005, France\\
$^{2}$Department of Mechanical Engineering and Program in Integrated
Applied Mathematics, University of New Hampshire, Durham, NH 03824, USA}

%%%% Subject entries to be placed here %%%%
\subject{mathematical physics, differential equations, fluid mechanics}

%%%% Keyword entries to be placed here %%%%
\keywords{multiple scale analysis, quasilinear systems, wave action, non-local energy transfers}

%%%% Insert corresponding author and its email address}
\corres{Guillaume Michel\\
\email{guillaume.michel@ens.fr}}

%%%% Abstract text to be placed here %%%%%%%%%%%%
\begin{abstract}
This article illustrates the application of multiple scales analysis to two archetypal quasilinear systems; i.e. to systems involving fast dynamical modes, called fluctuations, that are not directly influenced by fluctuation--fluctuation nonlinearities but nevertheless are strongly coupled to a slow variable whose evolution may be fully nonlinear.  In the first case, fast waves drive a slow, spatially-inhomogeneous evolution of their celerity field. Multiple scales analysis confirms that, although the energy $E$, the angular frequency $\omega$, and the modal structure of the waves evolve, the wave action $E/\omega$ is conserved in the absence of forcing and dissipation.  In the second system, the fast modes undergo an instability that is saturated through a feedback on the slow variable. A new multiscale analysis is developed to treat this case. The key technical point, confirmed by the analysis, is that the fluctuation energy and mode structure evolve slowly to ensure that the slow field remains in a state of near marginal stability. These two model systems appear to be generic, being representative of many if not all quasilinear systems. In each case, numerical simulations of both the full and reduced dynamical systems are performed to highlight the accuracy and efficiency of the multiple scales approach.  Python codes are provided as supplementary material.
\end{abstract}
%%%%%%%%%%%%%%%%%%%%%%%%%%%
\maketitle
%%%%%%%%%% Insert the texts which can accomdate on firstpage in the tag "fmtext" %%%%%

%\begin{fmtext}

\section{Introduction}
%%%% Insert A head here
Many physical, chemical, biological, and ecological systems evolve on disparate time scales.  By explicitly accounting for this scale separation, multiple scales analysis enables reduced equations governing the slow evolution of the comparably fast processes to be derived.  This flexible mathematical formalism has proven fruitful in a variety of contexts, not only enabling significant computational savings but also leading to the introduction of adiabatic invariants (e.g., wave action) and new governing equations (e.g., the nonlinear Schr\"odinger equation). In this article, we illustrate the derivation of reduced equations for  \textit{quasilinear} (QL) dynamical systems evolving on two distinct temporal scales. Slow--fast QL systems generically arise in the mathematical description of two broad classes of phenomena: (i) slowly-modulated waves, and (ii) instabilities that saturate through a feedback on the slow variable.  A primary theme of the present work is that there are fundamental differences in the mathematical analysis of these multi-scale QL systems, with the latter requiring the introduction of a new asymptotic formalism. 

QL systems have a characteristic mathematical structure comprising two sets of equations that govern the coupled evolution of slow variables, say $\bar{\mathbf{u}}$, and of fast fluctuations, say $\mathbf{u}'$. For instance, consider the Navier--Stokes equation for an incompressible and homogeneous fluid,
\begin{equation}
\frac{\partial \mathbf{u}}{\partial t} + \left( \mathbf{u} \cdot \mathbf{\nabla} \right) \mathbf{u} = -  \mathbf{\nabla} \left( \frac{P}{\rho}\right) + \nu \triangle \mathbf{u},\label{NS}
\end{equation}
where $\mathbf{u}$ is the velocity field, $P$ is the pressure, $\rho$ is the (constant) density, $\nu$ is the kinematic viscosity, and $\triangle$ is the Laplacian operator, and assume that the dynamics is characterized by fast waves coupled with a mean velocity field that evolves comparably slowly in time. To account for this scale separation, we introduce a Reynolds-like decomposition $\mathbf{u}= \bar{\mathbf{u}} + \mathbf{u}'$, where $\overline{\mathbf{u}'}=0$ and the bar refers to a (running) time average over many wave periods. With this expression, \eqref{NS} becomes
\begin{align} \label{NS_1}
&\frac{\partial \bar{\mathbf{u}}}{\partial t} + \left( \bar{\mathbf{u}} \cdot \mathbf{\nabla} \right) \bar{\mathbf{u}} + \overline{ \left( \mathbf{u}' \cdot \mathbf{\nabla} \right) \mathbf{u}'} = -  \mathbf{\nabla} \left( \frac{\overline{P}}{\rho}\right) + \nu \triangle \bar{\mathbf{u}},\\ \label{NS_2}
&\frac{\partial \mathbf{u}'}{\partial t} + \left( \bar{\mathbf{u}} \cdot \mathbf{\nabla} \right) \mathbf{u}' + \left( \mathbf{u}' \cdot \mathbf{\nabla} \right) \bar{\mathbf{u}} + \left( \mathbf{u}' \cdot \mathbf{\nabla} \right) \mathbf{u}' - \overline{ \left( \mathbf{u}' \cdot \mathbf{\nabla} \right) \mathbf{u}'}= -  \mathbf{\nabla} \left( \frac{P'}{\rho}\right) + \nu \triangle \mathbf{u}'.
\end{align}
The above set of equations is termed QL if the governing equations for the fluctuations $\mathbf{u}'$ do not include fluctuation--fluctuation nonlinearities; i.e., if the term $\left( \mathbf{u}' \cdot \mathbf{\nabla} \right) \mathbf{u}'$ [and, hence, $\overline{\left( \mathbf{u}' \cdot \mathbf{\nabla} \right) \mathbf{u}'}$] in \eqref{NS_2} is negligible.

Frequently, wave systems satisfy the QL approximation:  denoting the wavenumber of the waves by $k$,  the angular frequency by $\omega$, and the phase velocity (or ``celerity'') by $c$, then
\begin{equation}
\left \vert \frac{\left(\mathbf{u}' \cdot \mathbf{\nabla} \right)\mathbf{u}'}{\partial_t \mathbf{u}'} \right \vert \sim \frac{u' k}{\omega} \sim \frac{u'}{c}, \label{Mach}
\end{equation}
which is a (generalized) Mach number. If this number is small compared to unity, then \eqref{NS_2} reduces to QL form and the asymptotic methods introduced in this article can be applied. Since the nonlinear term $ \left( \mathbf{u}' \cdot \mathbf{\nabla} \right) \mathbf{u}'$ is neglected, quasilinearity also guarantees that the fluctuations cannot interact directly to generate new harmonics, thereby limiting the range of temporal scales that must be resolved in simulations. Nevertheless, the dynamics remains rich and the two-way nonlinear coupling between $\bar{\mathbf{u}}$ and $\mathbf{u}'$ is preserved: the cumulative effect of the fluctuation--fluctuation nonlinearities modifies the slow fields through the ``Reynolds stress'' divergence $\overline{ \left( \mathbf{u}' \cdot \mathbf{\nabla} \right) \mathbf{u}'}$ in \eqref{NS_1}, which in turn modifies the fluctuations through terms of the form $ \left( \bar{\mathbf{u}} \cdot \mathbf{\nabla} \right) \mathbf{u}'$ and $ \left(  \mathbf{u}' \cdot \mathbf{\nabla} \right) \bar{\mathbf{u}}$. These attractive attributes account for the increasing prevalence of QL models in the atmospheric, oceanic, and astrophysical sciences (although in these models the averaging operation frequently is  generalized to incorporate a spatial mean) and, more generally, in various branches of fluid mechanics, some examples of which are described below.

One well-documented QL phenomenon is the quasi-biennial oscillation of the winds in the lower equatorial stratosphere, which undergo reversals approximately every 14 months. This slowly-evolving flow results from a two-way coupling with internal gravity waves, which have a period of only a few tens of minutes\cite{Lindzen1968}: the waves drive a shear flow through their Reynolds stress divergence, and the shear flow in return refracts the waves.  Given the evident separation in temporal scales distinguishing the waves and the shear flow, along with the small Mach numbers characterizing the waves, these slow reversals can be captured in an idealized QL model\cite{Plumb1977,Plumb1978}. 
%As a second illustration, consider acoustic waves in a fluid exhibiting strong stable density inhomogeneities:  the nonlinear self-interaction of the waves can generate a mean Eulerian (``streaming'') flow that advects density inhomogeneities, thereby modifying the wave frequency, amplitude, and modal structure. 
As a second illustration, acoustic waves in a fluid exhibiting strong stable density inhomogeneities can nonlinearly interact to generate a mean Eulerian (``streaming'') flow that advects the density inhomogeneities, thereby modifying the wave frequency, amplitude, and modal structure.  This baroclinic acoustic streaming was first observed in high-intensity discharge lamps\cite{Dreeben2011}, with the wave period being $\sim 30~\mu \mathrm{s}$ and the streaming flow evolving on a timescale $\sim 0.1 \mathrm{s}$, and subsequently shown to be described accurately by a QL dynamical system \cite{Chini2014,AS_jfm, Karlsen2017}.

The application of multi-scale QL models is not restricted to dynamically-stable wave systems.  Indeed, the QL reduction also has proved surprisingly effective for the investigation of spatiotemporally-chaotic and even turbulent dynamical systems dominated by spectrally non-local energy transfers. In this latter scenario, e.g., for turbulent fluid flows, the fluctuation term $\mathbf{u}'$ represents ``eddies'' and quasilinearity is realized when $\mathbf{u}' \ll \bar{\mathbf{u}} $, hence $ \left( \mathbf{u}' \cdot \mathbf{\nabla} \right) \mathbf{u}' \ll  \left( \bar{\mathbf{u}} \cdot \mathbf{\nabla} \right) \mathbf{u}'$. In the atmospheres of the Earth and gas giants, for example, turbulent small-scale eddies are known to drive slowly-evolving zonal jets that in turn undergo shear instabilities and spawn small-scale eddies. Both the QL character and the time-scale separation manifest in this coupled eddy/jet system can be derived in the limit of small forcing and dissipation \cite{Bouchet2013}, thus providing a consistent framework to explain the spontaneous generation of these jets\cite{Tobias2013}. Similarly, in strongly (stably) stratified turbulence, anisotropic layers of horizontally-moving fluid (oriented orthogonally to the direction of the imposed density gradient) spontaneously emerge that, owing to their relative motion, are susceptible to small-scale instabilities. 
%(the resulting fluctuations being either wave- or eddy-like). 
A QL system has been derived in the asymptotic limit of strong stratification and shown to be capable of describing the dynamics of these anisotropic layers\cite{Chini_strat_turb}. Other examples of turbulent yet approximately quasilinear dynamical systems include strongly rotating (but weakly stratified) flows, such as open-ocean deep convection and high-latitude abyssal ocean currents\cite{Keith}, and rotating astrophysical disks in which magnetic fields are generated\cite{Squire}.

In this work, we describe the multiple scales analysis of two systems that are intended to be representative of the two broad classes of QL dynamics, involving waves and instabilities, respectively. These examples are sufficiently simple that the derivations are reasonably concise, allowing us to highlight the differences:  specifically, for QL systems exhibiting fast exponentially-growing instabilities rather than stable high-frequency wave motions, we show that the amplitudes of the ``most dangerous'' fluctuation modes are slaved to the slowly-evolving mean fields. The main novelty of our study is the new procedure we introduce to capture this slaving, which is not associated with the usual dissipative contraction to a slow manifold but rather with a marginal stability constraint.  Furthermore, to illustrate the utility of the multiple scales approach in each case, direct numerical simulations of the master equations are compared to numerical simulations of the reduced models we derive. The simulations are coded in Python, using the open-source framework Dedalus, which allows for an easy and efficient implementation of initial-value, boundary-value, and eigenvalue problems\cite{Dedalus}. Another attractive feature is that the user need only explicitly enter the equations, and the numerical scheme is then automatically generated.  For completeness, all of the associated source files are commented and provided as supplementary material.

\vfill{\eject}

\section{Modulated Waves}\label{MODULATEDWAVES}
\subsection{Governing equations}

The first system we analyze is a one-dimensional model of fast, linear waves that are coupled strongly to their celerity field.  For simplicity, only a bounded spatial domain is considered, although the analysis can be generalized to wave propagation in spatially-extended domains.  This system can be viewed as a generic model of wave/mean-flow interaction, crudely mimicking the phenomenology of, e.g., baroclinic acoustic streaming noted in the introduction. More generally, the analysis detailed in this section can be applied to any QL system in which the fluctuations are slowly-modulated waves that are \textit{not susceptible to a rapidly-amplifying instability}. Here, the celerity field $c$ and the wave field $\eta$ are chosen to satisfy the following governing equations (and initial conditions):
\begin{align} 
& \frac{\partial^2 c}{\partial t^2} = - (c-1)+ \frac{\partial^3 c}{\partial x^2 \partial t} - c \left( \frac{\partial \eta}{\partial x} \right)^2,& c(x,0)=1,~\frac{\partial c}{\partial t} (x,0)=0, \label{Mean_field}
\\ &\epsilon^2 \frac{\partial^2 \eta}{\partial t^2} = \frac{\partial}{\partial x} \left(c(x,t)^2 \frac{\partial \eta}{\partial x} \right),& \eta(x,0) = \frac{2}{\sqrt{L}} \cos \left( \frac{2 \pi x}{L} \right),~\frac{\partial \eta}{\partial t} (x,0)=0. \label{Wave_eq}
\end{align}
The spatial domain $x\in [0,L)$, and both $c$ and $\eta$ are $L$-periodic functions of $x$.
%complemented with periodic boundary conditions in a domain of length $L$. 
Equation~\eqref{Wave_eq} is the one-dimensional (1D) wave equation with inhomogeneous celerity, where $\epsilon^2$ is the square of a small parameter. (Note that all variables are dimensionless.) Equation~\eqref{Mean_field} describes the evolution of an initially uniform celerity distorted by the wave-induced Reynolds stress and subject to both restoring and dissipative processes. This set of equations conserves, in the absence of dissipation, the total energy $E_\mathrm{tot} = E_\mathrm{c} + E_\mathrm{w}$, where $E_\mathrm{c}$ and $E_\mathrm{w}$ are, respectively, the ``energy'' of the celerity field and of the waves, defined by
\begin{equation}
E_\mathrm{c} =\frac{1}{2} \int_0^L \left[ \left( \frac{\partial c}{\partial t} \right)^2 + (c-1)^2 \right] \mathrm{d}x,   ~~~E_\mathrm{w} = \frac{1}{2} \int_0^L \left[\epsilon^2 \left( \frac{\partial \eta}{\partial t} \right)^2 + c^2 \left( \frac{\partial \eta}{\partial x} \right)^2 \right]\mathrm{d}x. \label{Energy}
\end{equation}
For the given initial conditions, $c$ and $\eta$ are $\mathit{O}(1)$, implying the right-hand sides of \eqref{Mean_field} and \eqref{Wave_eq} also are of order unity.  Owing to the prefactor $\epsilon^2$ on the left-hand side of (\ref{Wave_eq}), which can be interpreted as a measure of fluctuation inertia, we expect the wave field $\eta$ to evolve on a much faster time scale than that characterizing the evolution of the celerity $c$.

\subsection{Leading-order equations}

The dynamics thus evolves on both a fast, $O(\epsilon)$, and a slow, $O(1)$, time scale. To exploit this temporal scale separation, we introduce a fast phase $\phi$ and a slow time $T$, where
\begin{equation}
\frac{\mathrm{d} \phi}{\mathrm{d}t } = \frac{\omega(T)}{\epsilon}, ~~~~~~~ T = t.
\end{equation}
Crucially, $\phi$ and $T$ are treated as \textit{independent variables} in the subsequent analysis.  The introduction of the phase $\phi$ through its non-constant time derivative captures the slow evolution of the wave angular frequency $\omega$ and is referred to as the WKBJ approximation\cite{Hinch}. Using the chain rule, $\partial_t \rightarrow (\omega(T) / \epsilon) \partial_\phi + \partial_T$, the governing equations become
\begin{align} 
&\frac{\omega^2}{\epsilon^2} \frac{\partial^2 c}{\partial \phi^2}  + \frac{1}{\epsilon} \left[2\omega \frac{\partial^2 c}{\partial \phi \partial T} +  \frac{\mathrm{d}\omega}{\mathrm{d}T }  \frac{\mathrm{\partial c}}{\partial\phi} \right] + \frac{\partial^2 c}{\partial T^2} = - (c-1) - c \left( \frac{\partial \eta}{\partial x} \right)^2  +  \frac{\omega}{\epsilon} \frac{\partial^3 c}{\partial x^2 \partial \phi} + \frac{\partial^3 c}{\partial x^2 \partial T} ,\label{Mean_field_mult}
\\
&\omega^2 \frac{\partial^2 \eta}{\partial \phi^2}  + \epsilon \left[2\omega \frac{\partial^2 \eta}{\partial \phi \partial T} +  \frac{\mathrm{d}\omega}{\mathrm{d}T }  \frac{\mathrm{\partial \eta}}{\partial\phi} \right] + \epsilon^2 \frac{\partial^2 \eta}{\partial T^2} =   \frac{\partial}{\partial x} \left( c^2 \frac{\partial \eta}{\partial x} \right). \label{Wave_eq_mult}
\end{align}
To proceed, $c$, $\eta$, and $\omega$ are expanded as asymptotic power series in $\epsilon$,
\begin{align} \label{expansion_eta}
&\eta(x,\phi, T) =\eta_0 (x,\phi, T) + \epsilon \eta_1 (x,\phi, T) + O(\epsilon^2),\\
& c(x,\phi,T) = c_0(x,\phi,T) + \epsilon c_1 (x,\phi,T) + O(\epsilon^2),\\
& \omega(T) = \omega_0(T) + \epsilon \omega_1(T) + O(\epsilon^2),
\end{align}
and substituted into (\ref{Mean_field_mult})--(\ref{Wave_eq_mult}).  Since $\epsilon$ is small but variable, the resulting equations then can be solved order by order. At $O(1/\epsilon^2)$, \eqref{Mean_field_mult} yields $\partial_\phi^2 c_0 =0$, and since linear growth of $c_0$ with $\phi$ would result in unbounded growth of $c_0$, we conclude that $c_0$ is a function of $x$ and $T$ only. This result confirms that at leading order $c$ is a field that evolves strictly on the slow time scale. With $\partial_\phi c_0 = 0$, \eqref{Mean_field_mult} at order $O(1/\epsilon)$ requires $\partial_\phi c_1=0$; i.e., $c_1$ also is a function only of $x$ and $T$. Finally, at $O(1)$, we obtain
\begin{equation}
\omega_0^2 \frac{\partial^2 c_2}{\partial \phi^2} + \frac{\partial^2 c_0}{\partial T^2} = - (c_0 - 1) - c_0 \left( \frac{\partial \eta_0}{\partial x} \right)^2 + \frac{\partial^3 c_0}{\partial x^2 \partial T}.\label{M1}
\end{equation}

We next introduce the fast average over $\phi$ (denoted with an overbar) of any function $f(x,\phi,T)$ bounded in $\phi$ :
\begin{equation}
\bar{f}(x,T) = {\lim_{n\to\infty}}\;\frac{1}{2n \pi} \int_{-n \pi}^{+n\pi} f(x,\phi,T) \mathrm{d}\phi,\;\;\;\;\mbox{for integer $n$}. \label{fast_scale_averaging}
\end{equation}
This averaging operation provides a suitable definition of the overbar used to define the Reynolds decomposition $\mathbf{u} = \bar{\mathbf{u}} + \mathbf{u}'$ in the introduction. 
%Applying this averaging procedure to \eqref{M1} {\textcolor{red}{eliminates}} the term involving $c_2$ and, since $\partial_\phi c_0=0$, yields
%%\textcolor{red}{a necessary condition for the sublinear growth of $c_2$:}
Applying this averaging procedure to \eqref{M1}, and recalling $\partial_\phi c_0=0$, yields a necessary condition for the sublinear growth of  $\partial_\phi c_2$:
\begin{equation}
\frac{\partial^2 c_0}{\partial T^2} = - (c_0 - 1) - c_0 \overline{\left( \frac{\partial \eta_0}{\partial x} \right)^2} + \frac{\partial^3 c_0}{\partial x^2 \partial T}. \label{WAVES_S1}
\end{equation}
In fact, it is readily confirmed using the resulting equation for $c_2$ [obtained by subtracting (\ref{WAVES_S1}) from (\ref{M1})] together with the expression for $\eta_0$ given below in (\ref{Waves_3}) that (\ref{WAVES_S1}) ensures the boundedness of $c_2$ on the slow time scale.

We now turn to the leading-order reduction of the fluctuation equation \eqref{Wave_eq_mult}:
\begin{equation}
- \omega_0(T)^2 \frac{\partial^2 \eta_0}{\partial \phi^2} +\frac{\partial}{\partial x} \left( c_0(x,T)^2 \frac{\partial \eta_0}{\partial x} \right)= 0. \label{WAVES_S2}
\end{equation}
Equations \eqref{WAVES_S1}--\eqref{WAVES_S2} comprise a two time-scale system, for which the fluctuation equation \eqref{WAVES_S2} does not involve nonlinear fluctuation--fluctuation terms (e.g., $\eta_0^2$) and thus is quasilinear. (Of course, for this system, no such nonlinearity was included in the master set of equations.) The key question that arises is how to accurately and efficiently integrate this two time-scale system?  We proceed by observing that the solution of \eqref{WAVES_S2} is of the form
\begin{align}
&\eta_0(x,\phi,T) = A(T) \hat{\eta}_0(x,T) \cos{\phi}, \label{Waves_3}
\end{align}
since $c_0$ does not depend on $\phi$ and noting that the solution component proportional to $\sin{\phi}$ vanishes as a result of the initial conditions \eqref{Wave_eq} on $\eta$.  Here, $A(T)$ is a slowly-evolving amplitude, and $\hat{\eta}_0$ is a function that characterizes the spatial structure of the wave mode. A normalization condition must be prescribed to make this decomposition unique; see \eqref{NORM}.  Crucially, at this stage of the analysis, the slow evolution of $A$ is unknown.  Indeed, the ansatz (\ref{Waves_3}) converts the \emph{initial-value} problem (\ref{WAVES_S2}) into a \emph{linear eigenvalue} problem that places no constraint on the modal amplitude.  Consequently, the reduced system is not closed on the slow time scale.  The objective of the remainder of the analysis is to derive an equation for $\mathrm{d}A/\mathrm{d}T$, which for this system is found by examining the equation for the correction field $\eta_1$.

\subsection{Inner product and adjoint operator\label{Sec_adjoint}}
Multiple scales analysis, in its most general form, requires certain notions from linear algebra that we introduce here. Substituting \eqref{Waves_3} into \eqref{WAVES_S2}, $\hat{\eta}_0$ is found to satisfy
\begin{equation}
\omega_0^2 \hat{\eta}_0 + \frac{\partial }{\partial x} \left( c_0^2 \frac{\partial \hat{\eta}_0}{\partial x} \right) = 0,~~~ \hat{\eta}_0(0,T) = \hat{\eta}_0(L,T), ~~~~~ \frac{\partial \hat{\eta}_0}{\partial x} (0,T) = \frac{\partial \hat{\eta}_0}{\partial x} (L,T).\label{S2_2}
\end{equation}
This eigenvalue problem is linear, and can be expressed more compactly as $\mathcal{L}  \hat{\eta}_0 = 0$, where the linear operator $\mathcal{L}=\omega_0^2 + \partial_x (c_0^2 \partial_{x})$, with periodic boundary conditions, acts on the function $\hat{\eta}_0$. For spatially-periodic and real-valued functions $f(x)$ and $g(x)$, we define the $L^2$ inner product
\begin{equation}
(f,g ) = \int_0^L f(x)g(x)~ \mathrm{d}x.\label{Inner_P}
\end{equation}
An inner product defines a norm, and we choose to set $(\hat{\eta}_0,\hat{\eta}_0)=1$; that is, 
\begin{equation}
\int_0^L  \hat{\eta}_0^2(x,T) \mathrm{d}x =1.\label{NORM}
\end{equation} 
This relation renders the decomposition \eqref{Waves_3} unique. Moreover, after integrating by parts and making use of the periodic boundary conditions, we obtain
\begin{equation}
( \mathcal{L} f,g ) =  \int_0^L  g(x) \left(\omega_0^2 + \partial_x (c_0^2 \partial_{x})  \right)f(x) \mathrm{d}x =   \int_0^L  f(x)  \left(\omega_0^2 + \partial_x (c_0^2 \partial_{x})  \right)g(x) \mathrm{d}x = (f , \mathcal{L}g).
\end{equation}
This equality reveals that $\mathcal{L}$ is \textit{self-adjoint}, a feature that, while not essential, simplifies the following analysis. For other boundary conditions (and/or other differential operators), the adjoint operator may not be equal to $\mathcal{L}$ but the following procedure can be appropriately modified \cite{Hinch}. As shown in the next subsection, computing the adjoint linear operator enables a \textit{solvability condition} to be imposed on the equation for the correction field $\eta_1$, which, in turn, ensures that the asymptotic expansion \eqref{expansion_eta} remains uniformly valid on the slow time scale $T$. This procedure generalizes the operation of ``removing resonant or secular forcing terms'' usually introduced in early lectures on multiple scales analysis\cite{Hinch}.
\subsection{Conservation of wave action\label{sec_WA}}
The slow temporal evolution of $A$ is obtained by collecting terms of order $O(\epsilon)$ in \eqref{Wave_eq_mult}:
\begin{equation}
-\omega_0^2 \frac{\partial^2 \eta_1}{\partial \phi^2} + \frac{\partial}{\partial x} \left( c_0^2 \frac{\partial \eta_1}{\partial x} \right) =  2 \omega_0 \frac{\partial^2 \eta_0}{\partial T \partial \phi} + \frac{\mathrm{d} \omega_0}{\mathrm{d} T} \frac{\partial \eta_0}{\partial \phi} + 2 \omega_0 \omega_1 \frac{\partial^2 \eta_0}{\partial \phi^2} - 2 \frac{\partial}{\partial x} \left( c_0c_1 \frac{\partial \eta_0}{\partial x} \right).
\label{S2}
\end{equation}
Note that the left-hand sides of \eqref{S2} and of  \eqref{WAVES_S2} involve the same linear operator. By setting $\eta_1(x,\phi,T) = \hat{\eta}_{1,\mathrm{c}}(x,T)\cos{\phi}+ \hat{\eta}_{1,\mathrm{s}}(x,T)\sin{\phi}$, \eqref{S2} yields
\begin{align} \label{EQ1}
&\mathcal{L} \hat{\eta}_{1,\mathrm{c}} = F_\mathrm{c},&F_\mathrm{c} = - 2 \omega_0 \omega_1 A \hat{\eta}_0 - 2A \frac{\partial}{\partial x} \left( c_0c_1 \frac{\partial \hat{\eta}_0}{\partial x} \right),\\ \label{EQ2}
&\mathcal{L} \hat{\eta}_{1,\mathrm{s}} = F_\mathrm{s},&F_\mathrm{s} = -2 \omega_0 \left( \frac{\mathrm{d}A}{\mathrm{d} T} \hat{\eta}_0 + A \frac{\delta \hat{\eta}_0}{\delta T} \right) - A \frac{\mathrm{d} \omega_0}{\mathrm{d}T}  \hat{\eta}_0.
\end{align}
The notation $\delta /\delta T$ is a short-hand for $(\partial_T c_0)\,\delta /\delta c_0$, where $ \delta /\delta c_0$ denotes \emph{functional differentiation} with respect to $c_0(x,T)$, as arises here because of the tight, two-way coupling between the waves and the slowly-evolving celerity field; specifically, the evolution of $c_0$ drives slow, $O(1)$ changes in the \textit{leading-order} wave eigenfunction $\hat{\eta}_0$, rendering the analysis non-standard. Nevertheless, we demonstrate below that the required functional derivatives can be evaluated explicitly, obviating the need for costly sensitivity computations.

Taking the inner product of \eqref{EQ1} with $\hat{\eta}_0$ and using the fact that $\mathcal{L}$ is self-adjoint, we obtain a solvability condition:
\begin{equation}
(F_\mathrm{c} , \hat{\eta}_0)= (\mathcal{L}\hat{\eta}_{1,\mathrm{c}}, \hat{\eta}_0) =(\hat{\eta}_{1,\mathrm{c}},  \mathcal{L}\hat{\eta}_0) =  ( \hat{\eta}_{1,\mathrm{c}},0) = 0; \label{Solv}
\end{equation}
i.e., the right-hand-side of (\ref{EQ1}) must be orthogonal to the null eigenvector of the (adjoint) linear operator.  Consequently, 
%Using \eqref{EQ1}, we find that
\begin{equation}
\omega_0 \omega_1 \int_0^L\hat{\eta}_0^2 \mathrm{d}x =- \int_0^L \hat{\eta}_0  \frac{\partial}{\partial x} \left( c_0c_1 \frac{\partial \hat{\eta}_0}{\partial x} \right) \mathrm{d}x = \int_0^L c_0c_1 \left( \frac{\partial\hat{\eta}_0}{\partial x}\right)^2   \mathrm{d}x.
\end{equation}
Using the normalization condition \eqref{NORM} then yields the following result:
\begin{equation}
\omega_1 = \frac{1}{\omega_0} \int_0^L c_0c_1 \left( \frac{\partial\hat{\eta}_0}{\partial x}\right)^2  \mathrm{d}x.\label{WMF_SC1}
\end{equation}
This expression for the angular frequency correction $\omega_1$, together with a relation for the evolution of the leading-order frequency $\mathrm{d}\omega_0 / \mathrm{d}T$ (see Sec.~\ref{Sec_domega}), is required only to close the system for the \textit{correction} fields $\omega_1$, $c_1$, $\eta_1$ arising at higher order, should they be desired.  In contrast, the solvability condition $(F_\mathrm{s},\hat{\eta}_0)=0$ yields a slow evolution equation for $A(T)$.  Specifically, we obtain
\begin{equation}
\left( 2 \omega_0 \frac{\mathrm{d}A}{\mathrm{d}T} + A \frac{\mathrm{d} \omega_0}{\mathrm{d} T}  \right) \int_0^L  \hat{\eta}_0^2 \mathrm{d}x  = - A \omega_0 \frac{\mathrm{d}}{\mathrm{d} T} \left( \int_0^L  \hat{\eta}_0^2 \mathrm{d}x \right) =0;
\end{equation}
the last equality is a consequence of the normalization \eqref{NORM}, following an interchange of the order of functional differentiation and integration. Thus, we obtain the \textit{amplitude equation}
\begin{equation}
2\omega_0\frac{\mathrm{d}A}{\mathrm{d}T}   + A \frac{\mathrm{d} \omega_0}{\partial T} = 0\quad \Rightarrow \quad \frac{\mathrm{d} }{\mathrm{d}T} \left( A ^2 \omega_0 \right)= 0.\label{WaveAc}
\end{equation}
The latter form of \eqref{WaveAc} reveals the existence of a conserved quantity in the limit of slow external processes (here, slow variation of the celerity $c$), termed an \textit{adiabatic invariant}. Given that the leading-order energy of the waves $E_0 = (\omega_0 A)^2/2$, as computed from \eqref{Energy}, the adiabatic invariant derived in \eqref{WaveAc} is the so-called \textit{wave action} $E_0 / \omega_0$. Conservation of wave action is a generic property of all slowly-modulated, non-dissipative linear waves\cite{Mcintyre}.

\subsection{Slow evolution of the angular frequency\label{Sec_domega}}

With the formalism introduced in Sec. \ref{Sec_adjoint}, it is possible to derive an additional equation that, although not necessary for this problem, is in fact crucial for predicting the slow dynamics of the system analyzed in Sec. \ref{sys2}.  Accordingly, this additional condition is introduced here to facilitate comparison between the wave and instability problems and is derived by differentiating the leading-order eigenvalue equation \eqref{S2_2} with respect to the slow time $T$:
\begin{equation}
\left[ \omega_0^2 + \frac{\partial}{\partial x} \left( c_0^2 \frac{\partial}{\partial x} \right) \right] \frac{\delta \hat{\eta}_0}{\delta T} = - 2 \left[ \omega_0 \frac{\mathrm{d} \omega_0}{\mathrm{d} T}  + \frac{\partial}{\partial x} \left( c_0 \frac{\partial c_0}{ \partial T}  \frac{\partial}{ \partial x} \right)\right] \hat{\eta}_0.\label{E3}
\end{equation}
This system is of the form of $\mathcal{L} (\delta \hat{\eta}_0 / \delta T) = G$, where $G$ is the right-hand side of \eqref{E3}. Thus, a solvability condition is obtained by taking the inner product of (\ref{E3}) with $\hat{\eta}_0$, yielding
\begin{equation}
\frac{\mathrm{d} \omega_0} {\mathrm{d} T} =-  \frac{1}{\omega_0} \int_0^L \hat{\eta}_0 \frac{\partial} {\partial x} \left( c_0 \frac{\partial c_0}{\partial T} \frac{ \partial \hat{\eta}_0}{\partial x} \right) \mathrm{d}x = \frac{1}{\omega_0} \int_0^L  \left( \frac{\partial  \hat{\eta}_0}{\partial x} \right)^2 c_0 \frac{\partial c_0}{\partial T} \mathrm{d}x.
\label{E4}
\end{equation}
This result provides an explicit evolution equation for the angular frequency of the waves. 

\subsection{Numerical implementation}

To assess the fidelity of the predicted slow dynamics to the actual system dynamics for small but finite values of $\epsilon$, direct numerical simulations (DNS) of the governing equations \eqref{Mean_field}--\eqref{Wave_eq} for various values of $\epsilon$ are compared to a numerical simulation of the reduced system \eqref{WAVES_S1}, \eqref{S2_2}, and \eqref{WaveAc} obtained from the multiple scales analysis. All simulations are performed in a domain of size $L=2 \pi$ with 32 grid points, use a second-order Runge-Kutta scheme, and output the time series of the energies $E_\mathrm{c}$ and $E_\mathrm{w}$ defined in \eqref{Energy}. Although not documented here, the convergence of each simulation with increasing spatiotemporal resolution has been confirmed.

The DNS are performed by solving the system  \eqref{Mean_field}--\eqref{Wave_eq} using a Fourier pseudospectral method, the timestep being set in accord with a CFL condition (e.g., $\mathrm{d}t \simeq 0.01$ for $\epsilon = 0.1$). The reduced model is discretized on a Chebyshev grid, a requirement of the Dedalus eigenvalue solver; note that only \eqref{WAVES_S1} is explicitly time-advanced. In this equation, the term involving the fluctuations can be expressed as
\begin{equation}
\overline{\left( \frac{\partial \eta_0}{\partial x}\right)^2}  = \frac{E_0(T)}{\omega_0(T)^2} \left( \frac{\partial \hat{\eta}_0}{\partial x} \right)^2,
\end{equation}
where $\omega_0(T)$ and $\hat{\eta}_0$ are obtained by numerically solving the eigenvalue problem \eqref{S2_2} at each slow timestep rather than time-advancing (\ref{WAVES_S2}) on the fast time scale, and the leading-order wave energy  $E_0(T) = E_0(0) \omega_0 (T) / \omega_0(0)$ owing to the conservation of wave action. For consistency with the initial conditions imposed in the DNS, we select $E_0(0)=\omega_0(0)=1$. Since the waves are not explicitly time-resolved, the timestep can be increased by an order of magnitude ($\mathrm{d}t = 0.1$) relative to that required for the DNS!

\begin{figure}[t!]
\centering
\includegraphics[width=4in]{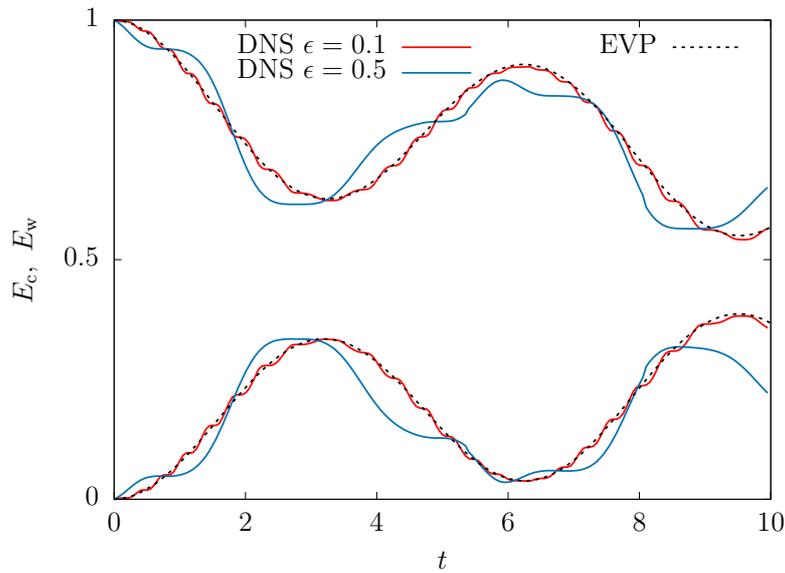}

\caption{Energy of the waves $E_\mathrm{w}$ ($E_\mathrm{w}(0)=1$) and of the celerity field $E_\mathrm{c}$ ($E_\mathrm{c}(0)=0$) obtained from direct numerical simulations (DNS) for $\epsilon=\{ 0.1,0.5\}$ and from the multiple scales model (derived in the limit $\epsilon\to 0$) in which the waves are computed via the solution of an eigenvalue problem (EVP) instead of being explicitly time-resolved.}
\label{Comparison}
\end{figure}

The time series of the energies are reported in Fig. \ref{Comparison}. Excellent agreement is achieved even for modest values of $\epsilon$,  and, for $\epsilon = 0.1$, the simulations of the reduced equations provide a quantitatively accurate representation of the energy exchanges between the waves and the external medium. Of course, for this unforced and dissipative system, the total energy decays toward zero and a rest state is eventually reached. As noted in the introduction, the associated source files for these simulations are commented and provided as supplementary material. The files also can be used to generate space--time diagrams of the celerity field, should they be desired.

\vfill{\eject}

\section{Unstable Fluctuations\label{sys2}}
\subsection{Governing equations}

The second example provides an illustration of a QL system in which the fluctuations would grow (or decay) exponentially fast in the absence of feedback on the slow variable. Nevertheless, we confirm that scale separation can be preserved as a result of the two-way coupling.  In this example, the slowly-evolving field $\Theta(x,t)$ and the fast fluctuation field $\eta(x,t)$ satisfy
\begin{align} \label{MF}
&\frac{\partial \Theta}{\partial t} = P - \nu \Theta - \eta^2, &\Theta(x,0) = -1,\\ \label{F}
& \epsilon \frac{\partial \eta}{\partial t} = \Theta \eta +  \frac{\partial^2 \eta}{\partial x^2} - \epsilon \eta^3, &\eta(x,0) =\mathrm{cos}\left(\frac{2 \pi x}{L} \right) , 
\end{align}
respectively, subject to periodic boundary conditions in a domain of spatial period $L$. It may be noted that the fluctuations satisfy a Ginzburg--Landau (GL) equation, e.g., describing the evolution of a non-uniform pattern amplitude $\eta(x,t)$, although neither this specification nor interpretation is a requirement of the formalism.  Moreover, unlike standard GL models, the ``distance'' to the instability threshold is controlled by the (slow) \emph{field variable} $\Theta$, which evolves according to \eqref{MF}, the terms on the right-hand side respectively representing an external forcing $P(x,t)>0$, a linear damping ($\nu > 0$ is a damping coefficient), and a quadratic feedback from the fluctuations. The small parameter $\epsilon$ controls the temporal scale separation. The coupled system (\ref{MF})--(\ref{F}) crudely mimics, for instance, the QL reduction of the equations governing Rayleigh--B\'{e}nard convection first proposed by Herring\cite{Herring}.

The coupling between the two fields $\Theta$ and $\eta$ is such that there exists an energy for this system, $E = E_\Theta + E_\eta$, where
\begin{equation}
E_\Theta = \frac{1}{2} \int_0^L \Theta^2~\mathrm{d}x,~~~~~~~E_\eta = \frac{\epsilon}{2} \int_0^L \eta^2~\mathrm{d}x.
\end{equation}
It transpires that at leading order in $\epsilon$ the nonlinear term in \eqref{F} is negligible and, thus, the total energy $E$ is conserved in the absence of forcing ($P=0$) and dissipation (requiring both $\nu=0$ in \eqref{MF} and zero diffusion in \eqref{F}).  In the absence of the nonlinear term in (\ref{F}), the system (\ref{MF})--(\ref{F}) becomes quasilinear.

\subsection{Leading-order equations}
As for the slowly-modulated wave system considered in Sec.~\ref{MODULATEDWAVES}, the small parameter $\epsilon$ motivates the introduction of two scales $\phi$ and $T$, treated hereafter as independent variables and defined according to
\begin{equation}
\frac{\mathrm{d}\phi}{\mathrm{d}t} = \frac{\sigma(T)}{ \epsilon}, ~~~~~~~~ T = t.\label{WKBJex2}
\end{equation}
The notation $\sigma$ rather than $\omega$ is used in this example to signify that the fluctuations may grow (or decay) exponentially rather than oscillate rapidly. Like $\omega$, $\sigma$ is defined to be an eigenvalue of a certain linear operator.  A crucial distinction, however, is that while $\omega$ is the slowly-varying angular frequency of \emph{any} one of a countable infinity of wave modes (see Sec.~\ref{MODULATEDWAVES}), $\sigma$ is the slowly-evolving instantaneous growth rate of the \emph{most unstable} (or least stable) fluctuation mode. %that will be defined formally as the largest (real) eigenvalue of specific eigenvalue problem.}}
%\textcolor{red}{Moreover, rather than tracking any given wave mode of slowly-evolving angular frequency $\omega(T)$ as in Sec.~\ref{MODULATEDWAVES}, here we are obliged to define $\sigma$ to be the slowly-evolving instantaneous growth rate of the \emph{most unstable} (or least stable) fluctuation mode.}
%\textcolor{red}{Instead of tracking any given wave mode of slowly evolving angular frequency, we shall see that for this problem $\sigma$ will correspond to the (slowly evolving) instantaneous growth rate of the most unstable fluctuation mode.} 

Using (\ref{WKBJex2}), the governing equations \eqref{MF}--\eqref{F} become
\begin{align}\label{MF_1}
&\frac{\sigma}{\epsilon} \frac{\partial \Theta}{\partial \phi}+ \frac{\partial \Theta}{\partial T} = P - \nu \Theta - \eta^2,\\ \label{F_1}
&\frac{\sigma}{\epsilon} \frac{\partial \eta}{\partial \phi} +  \frac{\partial \eta}{\partial T} =\frac{\Theta \eta}{\epsilon} + \frac{1}{\epsilon} \frac{\partial^2 \eta}{\partial x^2} - \eta^3.
\end{align}
This set of equations is solved order by order in $\epsilon$ after substituting the following asymptotic expansions for $\sigma$, $\Theta$, and $\eta$:
\begin{align}
&\sigma(T)= \sigma_0(T) + \epsilon \sigma_1(T) + O (\epsilon^2),\\
&\Theta(x,\phi ,T) = \Theta_0 (x,\phi, T) + \epsilon \Theta_1 (x,\phi,T) + O(\epsilon^2),\\
&\eta(x,\phi ,T) = \eta_0 (x,\phi, T) + \epsilon \eta_1 (x,\phi,T) + O(\epsilon^2).\label{expansion}
\end{align}
Equation \eqref{MF_1} at $O(1/\epsilon)$ yields $\partial_\phi \Theta_0=0$; i.e., $\Theta_0$ is a function strictly of $x$ and $T$. At $O(1)$, the fast average introduced in \eqref{fast_scale_averaging} removes the term proportional to $\partial_\phi\Theta_1$, yielding
\begin{equation}
\frac{\partial \Theta_0}{\partial T } = P - \nu \Theta_0 - \overline{\eta_0^2}. \label{T0}
\end{equation}
The leading-order fluctuation equation [i.e. Eq.~\eqref{F_1} at $\mathit{O}(1/\epsilon)$] is 
%sufficient to close the system: 
\begin{equation}
\sigma_0 \frac{\partial \eta_0}{\partial \phi} = \Theta_0 \eta_0 + \frac{\partial^2 \eta_0}{\partial x^2}. \label{T1}
\end{equation}
Equations  \eqref{T0}--\eqref{T1} comprise a two time-scale system that (as for the modulated waves example) can be closed on the \emph{fast} time scale upon making the non-asymptotic replacement $\partial_T\to(\sigma/\epsilon)\partial_\phi$.  This system is also quasilinear: fluctuation--fluctuation nonlinearities are of higher order and have been consistently neglected in the fluctuation equation~(3.11). Typically, the nonlinearity in (3.2), the equation from which (3.11) is derived, is crucial for the saturation of amplifying solutions that obey Ginzburg--Landau dynamics; its systematic omission in (3.11) implies that a distinct saturation mechanism is involved here.
  
Once again, the question that arises is how to accurately and efficiently integrate this slow--fast QL system when two formally independent time scales are retained?  The combination of temporal scale separation and quasilinearity enables a modal solution of \eqref{T1} to be expressed as
%Once again, the question that arises is how to accurately and efficiently integrate this two time-scale system?  The combination of temporal scale separation and quasilinearity enables {\textcolor{red}{a modal}} solution of \eqref{T1}  -- when two formally independent time scales are retained -- to be expressed as
\begin{equation}\label{solution_EVP}
\eta_0(x,\phi,T) = A(T) \hat{\eta}_0(x,T) e^{\phi},
\end{equation}
where $A(T)$ is the slowly-varying modal amplitude, and $\hat{\eta}_0$ describes the spatial structure of the fluctuation mode.  Note that $A$, $\hat{\eta}_0$, and $\phi$ are real-valued and that a suitable normalization condition on $\hat{\eta}_0$ is required to render the decomposition (\ref{solution_EVP}) unique; see \eqref{NORM_INST}.  Upon substituting (\ref{solution_EVP}) into (\ref{T1}), $\sigma_0$ is identified as the maximum eigenvalue of the operator $(\Theta_0 + \partial_x^2)$.
%that is, in the limit of strict scale separation, the linear, homogeneous, and autonomous initial-value problem (\ref{T1}) is best re-interpreted as an eigenvalue problem.
The leading-order fluctuation equation (\ref{T1}) seemingly places no constraint on the evolution of $A(T)$, confirming that the system (\ref{T0})--(\ref{T1}) is \emph{not} closed on the slow time scale.  
%\textcolor{red}{Consequently, the maximum instantaneous growth rate $\sigma_0$ corresponds precisely to the maximal eigenvalue of the operator $(\Theta_0 + \partial_{xx})$.}

Of course, the potential exponential growth (or decay) of the leading-order fluctuation field $\eta_0$ on the fast time scale would, if not properly addressed, break the posited scalings, as evidenced by the force acting on the slow field $\Theta_0$:
\begin{equation}
\overline{\eta_0^2} = \lim_{n\to\infty}\;\frac{A^2 \hat{\eta}_0^2}{2 n \pi} \int_{-n\pi}^{+n \pi} e^{2\phi} \mathrm{d}\phi .\label{FORCE}
\end{equation}
Note that, as for the fast temporal average introduced in \eqref{fast_scale_averaging}, the fast time variable 
%(i.e., roughly $T/\epsilon$) 
ranges from negative to positive infinity (rather than from, e.g., zero to positive infinity) while the slow time $T$ is fixed.  To ensure the convergence of this fast-time average for all $x$, one of the following two conditions must be satisfied: (i) $A(T) = 0$ or (ii) $\sigma(T)=\epsilon \mathrm{d}\phi/\mathrm{d}t = 0$ (implying the integrand in (\ref{FORCE}) is constant).
%[{\textcolor{red}{i.e., there is no fast-time evolution}}].
%[i.e., $\phi(t)$ is constant]. 
These conditions thus require $A(T)=0$ if the maximum instantaneous growth rate $\sigma(T)$ is non-zero, a natural consequence of exponentially-fast damping for $\sigma <0$ and a requirement, were $\sigma>0$, to preclude the blow-up of the fluctuations (and, hence, of the force (\ref{FORCE})) on the fast time scale.  With $A\equiv 0$, the leading-order dynamics reduces to \eqref{T0} with $\overline{\eta_0^2}=0$.  Conversely, the fluctuations can have finite amplitude \textit{only} if $\sigma=0$; i.e, only if the fluctuations are \textit{marginally stable}.  Although $\sigma$ initially need not equal zero, it continuously evolves under the forcing of the slow field and may eventually pass through zero.  Heuristically, the persistence of one or more zero eigenvalues may be understood by recognizing that (i)~$\Theta_0$ is not an arbitrary function of $x$ owing to the feedback from the fluctuations, and (ii)~scale-separated quasilinear systems with unstable fast fluctuations must self-tune to a state of near marginal stability.  Indeed, as we demonstrate through this example, it is this regime that characterizes the long-term dynamics of the forced system. 

Thus, if the largest eigenvalue $\sigma_0$ differs from zero, then the prescription is made that the modal amplitude $A = 0$.  If, however, $\sigma_0=0$, then the coupled system \eqref{T0}--\eqref{T1} is not closed on the slow time scale:
%On the contrary, since  \eqref{T1} is a linear homogeneous eigenvalue problem, for $\sigma_0=0$ the coupled system  \eqref{T0}--\eqref{T1} is not closed on the slow time scale:}   
a slow evolution equation for $A(T)$ is required.  In the next two subsections, we show how to derive such an equation.  Although we enforce $\sigma=0$ in the following analysis, we formally retain $\sigma_0$ explicitly in the derivation for clarity and generality of exposition.  Since $\phi$ is constant for $\sigma=0$, the amplitude $A$ can be re-defined so that the leading-order fluctuation field
\begin{equation}
\eta_0(x,\phi,T)=A(T) \hat{\eta}_0(x,T)
\end{equation}
and, accordingly, the fast-time average $\overline{\eta_0^2}=A^2\hat{\eta}_0^2$.
The leading-order system therefore becomes
\begin{align}
&\left(-  \sigma_0 + \Theta_0 \right) \hat{\eta}_0 + \label{EVP_instab} \frac{\partial^2 \hat{\eta}_0}{\partial x^2} = 0,\\ 
& \frac{\partial \Theta_0}{\partial T } = P - \nu \Theta_0 - A(T)^2 \hat{\eta}_0^2.\label{MF_Instab}
\end{align}
%\textcolor{red}{Consequently, the maximum instantaneous growth rate $\sigma_0$ corresponds precisely to the maximal eigenvalue of the operator $(\Theta_0 + \partial_{xx})$.}

\subsection{Solvability condition}
As in Sec.~2.\ref{Sec_adjoint}, the eigenvalue problem \eqref{EVP_instab} can be cast into the form $\mathcal{L} \hat{\eta}_0= 0$, now with $\mathcal{L} = -\sigma_0 + \Theta_0 + \partial^2_{x}$. As a result of periodicity, the linear operator $\mathcal{L}$ is self-adjoint with the inner product defined in \eqref{Inner_P}; i.e., 
\begin{equation}
(\mathcal{L}f,g) = \int_0^L g(x) (-\sigma_0 + \Theta_0 + \partial^2_{x}) f(x) \mathrm{d}x=\int_0^L f(x) (-\sigma_0 + \Theta_0 + \partial_{xx}) g(x) \mathrm{d}x=(f, \mathcal{L}g).
\end{equation}
This inner product is used to disentangle $A$ and $\hat{\eta}_0$ by imposing $(\hat{\eta}_0,\hat{\eta}_0)=1$; that is,
\begin{equation}
\int_0^L \hat{\eta}_0^2 \mathrm{d}x = 1.\label{NORM_INST}
\end{equation}
Guided by the analysis of the wave system, we first attempt to derive a slow-evolution equation for the amplitude $A$ through the ``usual procedure'' introduced in Sec. \ref{MODULATEDWAVES} \ref{sec_WA}; namely, by ensuring that the higher-order fluctuation equations are solvable.  Specifically, collecting terms at $O(1)$ in \eqref{F_1} leads to
\begin{equation}
\mathcal{L}\eta_1 = F,~~~~~~~~F =  \frac{\partial \eta_0}{\partial T} - \Theta_1 \eta_0  + \eta_0^3 = \left( \frac{\mathrm{d}A}{\mathrm{d}T} -A \Theta_1 + A^3 \hat{\eta}_0^2 \right) \hat{\eta}_0 + A \frac{\delta \hat{\eta}_0}{\delta T},\label{eta1eqn}
\end{equation}
where $\delta/\delta T=(\partial_T\Theta_0)\delta/\delta\Theta_0$.
Forming the inner product of (\ref{eta1eqn}) with $\hat{\eta}_0$, we find that $(\mathcal{L}\eta_1 ,\hat{\eta}_0) = (\eta_1 ,\mathcal{L}\hat{\eta}_0)=0$ and, hence, that $ (F ,\hat{\eta}_0)=0$.  Noting that the integral involving the functional derivative in $ (F ,\hat{\eta}_0)$ vanishes as a result of the normalization condition, we obtain
\begin{equation}
\frac{1}{A}\frac{\mathrm{d} A}{\mathrm{d} T} + A^2 \int_0^L \hat{\eta}_0^4 \mathrm{d}x =  \int_0^L \Theta_1 \hat{\eta}_0^2 \mathrm{d}x.\label{THETA1}
\end{equation}

In the wave problem, two solvability conditions emerge from the equation for $\eta_1$: a closure, (\ref{WMF_SC1}), required to evolve the dynamics of the correction fields $\omega_1$, $\eta_1$, and $c_1$ and an amplitude equation, \eqref{WaveAc}, prescribing $\mathrm{d}A/\mathrm{d}T$ as a function of the leading-order variable $\omega_0$. In contrast, in the present problem, only the single constraint \eqref{THETA1}, relating  the time evolution of $A$ to the \textit{higher-order} field $\Theta_1$, is obtained.  To procure a closed set of equations, it is natural to attempt to include $\Theta_1$ in the set of unknown fields.  An equation for $\Theta_1$ can be derived from \eqref{MF_1} at $O(\epsilon)$, yielding $\partial_T \Theta_1$ as a function of $\eta_0$ and, disappointingly, of $\eta_1$, another unknown variable.  Thus, similarly to \eqref{WMF_SC1}, \eqref{THETA1} is merely a constraint required to obtain the dynamics of the correction fields $\eta_1$ and $\Theta_1$.

\subsection{Slow evolution of the growth rate}
In this example, a suitable constraint on the amplitude $A$ can be derived by employing an operation analogous to that performed (solely for illustrative purposes) in Sec.~2\ref{Sec_domega}. Specifically, we differentiate the leading-order fluctuation equation \eqref{EVP_instab} with respect to $T$, yielding
\begin{equation}
\mathcal{L} \frac{\delta \hat{\eta}_0}{\delta T} = \hat{\eta}_0 \left(\frac{\mathrm{d} \sigma_0}{\mathrm{d} T} -  \frac{\partial \Theta_0}{\partial T}  \right).\label{evalueDiffT}
\end{equation}
%where the notation $\delta / \delta T$ refers to $(\partial \Theta_0 / \partial T) \delta / \delta \Theta_0$. 
A solvability condition is obtained by taking the inner product of (\ref{evalueDiffT}) with $\hat{\eta}_0$, again utilizing the normalization condition on $\hat{\eta}_0$; \emph{viz.},
\begin{equation}
\frac{\mathrm{d} \sigma_0}{\mathrm{d} T} = \int_0^L   \hat{\eta}_0^2  \left(\frac{\partial \Theta_0}{\partial T}\right)  \mathrm{d}x.\label{dt_sigma_orig}
\end{equation}
Upon substituting \eqref{MF_Instab} into (\ref{dt_sigma_orig}), this solvability condition reduces to 
\begin{equation}
\frac{\mathrm{d} \sigma_0}{\mathrm{d} T} = \int_0^L     \left( P - \nu \Theta_0 - A(T)^2 \hat{\eta}_0^2\right) \hat{\eta}_0^2 ~\mathrm{d}x \equiv \alpha - \beta A(T)^2,\label{dt_sigma}
\end{equation} 
where the real coefficients $\alpha$ and $\beta$ are defined by
\begin{equation}
\alpha = \int_0^L (P - \nu \Theta_0) \hat{\eta}_0^2 ~\mathrm{d}x ,~~~~~~~ \beta =\int_0^L \hat{\eta}_0^4 ~\mathrm{d}x.\label{alpha_beta}
\end{equation}

Equation \eqref{dt_sigma} prescribes the slow evolution of $\sigma_0$ as a function of the variables $\Theta_0$ and $\hat{\eta}_0$ and, crucially, of the fluctuation amplitude $A$. Since, for non-zero $A$, $\sigma_0$ must be zero, the consistency of the multiple scales expansion excludes strictly positive values of $\mathrm{d}\sigma_0/\mathrm{d}T$, which would immediately lead to positive growth rates; i.e., to exponentially-growing fluctuations. This scenario arises only when $\alpha>0$ and $\sigma_0=0$, in which case the amplitude $A(T)$ must be set to enforce $\mathrm{d}\sigma_0 / \mathrm{d}T=0$:
\begin{equation}
A(T) = \sqrt{\frac{\alpha}{\beta}}~~~~~~ \mathrm{if}~\sigma_0=0~\mathrm{and}~\alpha >0. \label{A}
\end{equation}
Equation \eqref{A} describes the effectively instantaneous (on the slow time scale) saturation of the instability via the feedback of the fluctuation field $\eta_0$ on the slow variable $\Theta_0$.

\subsection{Numerical implementation}

To demonstrate the efficacy of the proposed multi-scale algorithm, DNS of the governing equations \eqref{MF}--\eqref{F} are compared to simulations of the multiple scales reduction given by \eqref{EVP_instab}, \eqref{MF_Instab}, and \eqref{A}. All simulations are performed in a domain of size $L=2 \pi$ that is  discretized with 32 grid points.  The damping coefficient $\nu = 1$, and the external forcing
\begin{equation}
P = 1+\frac{1}{2}\bigg(\cos(t) \cos(x)+ \sin(0.6t) \cos(2x)\bigg).
\end{equation}
A second-order Runge--Kutta time-stepping scheme is used with fixed time step $\mathrm{d}t=0.01$ for the reduced model and $\mathrm{d}t=0.005$ for the DNS. In the DNS, the governing equations \eqref{MF}--\eqref{F} are solved using a Fourier pseudospectral method for $\epsilon= \{0.1, 0.01 \}$, while the reduced equations (valid as $\epsilon\to 0$) are discretized using Chebyshev polynomials on a non-uniformly spaced grid.  For simulations of the reduced system, only the mean equation \eqref{MF_Instab} is explicitly evolved in time (recall that $T=t$).  At every slow timestep, however, the eigenvalue problem \eqref{EVP_instab} is solved to obtain $\sigma_0$ and $\hat{\eta}_0$.  Finally, the amplitude of the fluctuations $A$ is set according to
\begin{equation}
A(T) = \left\{ 
\begin{array}{l l}
  0 & \quad \text{if}~ \sigma_0 < 0   ~\mathrm{or}~ \alpha <0,\\
  \sqrt{\alpha / \beta} & \quad \text{otherwise},\\ 
\end{array} 
\right\}
\end{equation}
where the slowly-varying coefficients $\alpha$ and $\beta$ have been defined in \eqref{alpha_beta}.

\begin{figure}[t!]
\centering
\includegraphics[width=4in]{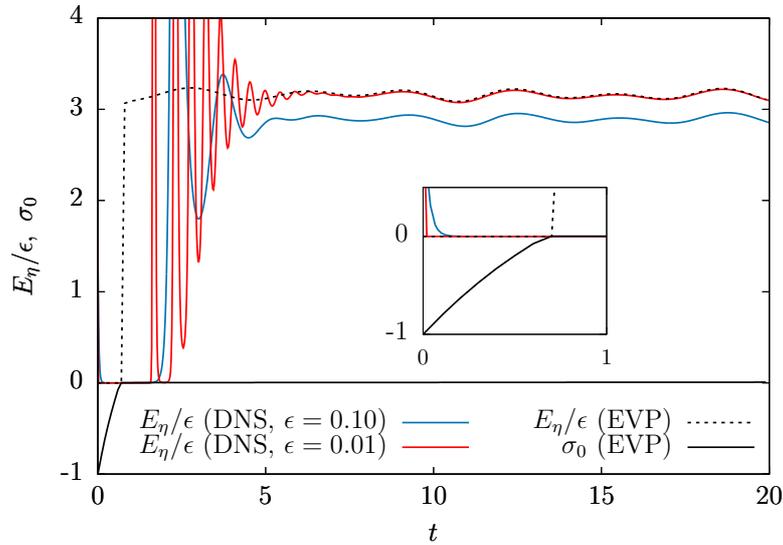}

\caption{Normalized energy of the fluctuations $E_\eta/\epsilon $ computed from direct numerical simulations (DNS) for $\epsilon=\{ 0.1,0.01\}$ and from a novel multiple scales algorithm in which the fluctuations are not time-evolved but computed from the solution to an eigenvalue problem (EVP). The maximum instantaneous growth rate, $\sigma_0$, obtained from the solution to the EVP is also reported.}
\label{Comparison_nowave}
\end{figure}

As evident in Fig. \ref{Comparison_nowave}, this procedure prevents $\sigma_0$ from becoming positive, thereby constraining the system to evolve along a marginal stability manifold. In practice, discretization and round-off errors 
%incurred by our numerical approximations 
preclude $\sigma_0$ from identically equalling zero.  Accordingly, for the results reported here, a tolerance of 0.01 has been enforced: $0\le\sigma_0\le 0.01$. (Smaller tolerances can be achieved by decreasing the time step $\mathrm{d}t$).
% \textcolor{red}{Numerically, $\sigma_0$ cannot strictly equal zero but remains smaller than 0.01 (smaller values can be achieved by decreasing $\mathrm{d}t$).}  
 The time series of the normalized fluctuation energy (half the norm of $\eta$),
\begin{equation}
\frac{E_\eta}{\epsilon} = \int_0^L \frac{\eta^2}{2} \mathrm{d}x,
\end{equation}
is also plotted in this figure.  (Although not plotted here, space--time diagrams of both $\eta$ and $\Theta$ can be generated using the source files provided as supplementary material.)  Excellent agreement between the DNS and the reduced model is observed for$\epsilon \simeq 0.01$ \textit{after} a transient that persists for a few slow time units. This ``bursting'' regime, evident in the DNS, is very sensitive to the initialization of the fluctuations, because the amplitude of the fluctuations is extremely small when positive growth rates are first attained.  (Recall that, before this instant, the fluctuations experience approximately exponential decay.)  This transient is absent from the reduced model, since the energy of the fluctuations is instantaneously adjusted to a finite value once a state with zero growth rate is reached. 

\subsection{Additional comments}

The multiple scales analysis described in this section applies to QL systems in which fluctuations have the potential to be amplified via a linear instability mechanism. These systems may have other attributes that warrant further discussion, as detailed below.

%\subsubsection{Unstable waves}
\subsubsection{Oscillatory instabilities}
%The key requirement of the algorithm -- tuning the amplitude of the fluctuations to prevent their exponential growth -- results from the possibility that the \emph{real} part of the growth rate $\sigma$ could become positive, regardless of the value and sign of the imaginary part. For the system analyzed in this section, the imaginary part of the growth rate is strictly zero; i.e., the instability is stationary.  Nevertheless, a similar procedure can be applied when a bifurcation gives rise to growing wave modes.  These dynamics would, for instance, be admitted by the following modified system:
The key requirement of the algorithm -- tuning the amplitude of the fluctuations to prevent their exponential growth -- results from the possibility that the \emph{real} part of the growth rate $\sigma$ could become positive, regardless of the value and sign of the imaginary part. For the system analyzed in this section, the imaginary part of the growth rate is strictly zero; i.e., the instability is stationary.  Nevertheless, a similar procedure can be applied when a bifurcation gives rise to an oscillatory instability, as, for instance, can occur in the following modified system:
\begin{align}
&\frac{\partial \Theta}{\partial t} = P - \nu \Theta + \nu \frac{\partial^2 \Theta}{\partial x^2} - \epsilon^2 \left( \frac{\partial \eta}{\partial t} \right)^2, &\Theta(x,0) = -1,\\ \label{F_unst_wave}
& \epsilon^2 \frac{\partial^2 \eta}{\partial t^2}  = \epsilon \Theta \frac{\partial \eta}{\partial t} +  \frac{\partial^2 \eta}{\partial x^2}, & \eta(x,0) = \mathrm{cos}\left( \frac{2 \pi x}{L} \right),  ~\frac{\partial \eta}{\partial t}(x,0) = 0  .
\end{align}
In this example, \eqref{F_unst_wave} describes fast oscillatory motions (waves) of frequency $O(1/\epsilon)$ that are either damped or exponentially amplified by the slow field $\Theta$.  Generally, the growth rate $\sigma$ is complex, but the amplitude of the fluctuations must be set, in the multiple scales analysis, by the requirement that the real part of $\sigma$ should not increase once a state of marginal stability is reached.  
%In this example, \eqref{F_unst_wave} describes fast waves of frequency $O(1/\epsilon)$ that are either damped or exponentially amplified by the slow field $\Theta$.  Generally, the growth rate $\sigma$ is complex, and the amplitude of the fluctuations must be set, in the multiple scales analysis, by the requirement that the real part of $\sigma$ should not increase once a state of marginal stability is reached.  

\subsubsection{Stabilizing nature of the fluctuation-induced feedback}
Omitting the subscript `0', the time-derivative of the (real) growth rate $\sigma$ obtained in \eqref{dt_sigma} is given by
\begin{equation}
\frac{\mathrm{d} \sigma}{\mathrm{d}T} = \alpha - \beta A(T)^2, 
\end{equation} 
where, again, $T$ is the slow time variable, $A$ is the amplitude of the fluctuations, and $\alpha$ and $\beta$ are two real parameters. The multiple scales approach developed in this section only applies if $\beta >0$, regardless of the value of $\alpha$. (If $\alpha >0$ and $\sigma=0$, then $A = \sqrt{\alpha / \beta}$; otherwise $A = 0$). That is, the fluctuations must provide a \textit{restoring} force, not drive the slow field toward an even more unstable state.  Although $\beta >0$ in the model analyzed here, see \eqref{alpha_beta}, this property is not generic; a slight modification of the set of equations \eqref{MF}--\eqref{F} provides a counter-example:
\begin{align} \label{MF_2}
&\frac{\partial \Theta}{\partial t} = P - \nu \Theta - \eta^2 e^{-\Theta^2},\\ \label{F_2}
& \epsilon \frac{\partial \eta}{\partial t} = \Theta \eta e^{-\Theta^2} +  \frac{\partial^2 \eta}{\partial x^2} - \epsilon \eta^3.
\end{align}
An analysis similar to that applied to (\ref{MF})--(\ref{F}) confirms that, for small values of $\epsilon$, this system is of multi-scale QL form. The corresponding expression for $\beta$, however, is modified,
\begin{equation}
 \beta =\int_0^L \hat{\eta}_0^4 ~\mathrm{d}x \Rightarrow  \beta =\int_0^L \hat{\eta}_0^4 (1-2 \Theta_0^2)e^{-2 \Theta_0^2} ~\mathrm{d}x,
\end{equation}
so that, crucially, $\beta$ is no longer sign-definite. In the scenario $\sigma=0$, $\alpha>0$, and $\beta<0$, the fluctuations cannot prevent positive values of the growth rate from being realized, thereby invalidating the posited asymptotic scalings.  In fact, the resulting exponential growth would be saturated by the nonlinear term in \eqref{F_2} by a physical process not captured by the QL system. Nevertheless, asymptotic methods still can be applied to this regime via a suitable rescaling of the unknowns:
\begin{align}
&\sigma(T)= \sigma_0(T) + \epsilon \sigma_1(T) + O (\epsilon^2)\\
&\Theta(x,\phi ,T) = \Theta_0 (x,\phi, T) + \epsilon \Theta_1 (x,\phi,T) + O(\epsilon^2),\\
&\eta(x,\phi ,T) = \frac{\eta_0 (x,\phi, T)}{\sqrt{\epsilon}} + \sqrt{\epsilon} \eta_1 (x,\phi,T) + O(\epsilon \sqrt{\epsilon}).
\end{align}
With these expansions, the following set of equations is obtained at leading order in $\epsilon$:
 \begin{align} \label{MF_3}
&\sigma_0\frac{\partial \Theta_0}{\partial \phi} = - \eta_0^2 e^{-\Theta_0^2},\\ \label{F_3}
& \sigma_0\frac{\partial \eta_0}{\partial \phi} = \Theta_0 \eta_0 e^{-\Theta_0^2} +\frac{\partial^2 \eta_0}{\partial x^2}- \eta_0^3,
\end{align}
where the leading-order growth rate $\sigma_0$ is still defined by the \textit{linear} eigenvalue problem obtained from \eqref{F_3} without the nonlinear term.
Thus, in this regime, the system evolves strictly on the fast time scale, with the dynamics taking the form of punctuated bursting.  The asymptotic analysis also reveals that, in this regime, the slow external forcing $P$ and damping --$\nu \Theta_0$ do not contribute to the leading-order dynamics. Numerically, both \eqref{MF_3} and \eqref{F_3} must be co-evolved on the fast time scale until $\sigma_0$ reaches zero again, and the dynamics relaxes to a slow manifold describable by the asymptotic QL reduction.  

\subsubsection{Degeneracy of the marginal eigenvalue}
Our last point concerns the assumption, implicitly made throughout this section, that the eigenvalue problem \eqref{EVP_instab} has non-degenerate eigenvalues, at least for the eigenvalue having the largest growth rate (i.e., $\sigma = 0$ is a simple eigenvalue), in accord with the ansatz \eqref{solution_EVP}. In fact, the non-degeneracy of the leading eigenvalue can be demonstrated for this particular eigenvalue problem.  Consider \eqref{EVP_instab}, and rewrite the variables as follows, 
\begin{equation}
\hat{\eta}_0 \rightarrow \Psi, ~~~~~\sigma_0 \rightarrow - E, ~~~~~~ \Theta_0 \rightarrow - V.
\end{equation}
Equation \eqref{EVP_instab} then becomes 
\begin{equation}
E \Psi = H \Psi, ~~~~~~H = - \frac{\partial^2}{\partial x^2} +V,
\end{equation}
subject to periodic boundary conditions.  Thus, the eigenvector of largest growth rate $\sigma_0$ formally is equivalent to the ground state of the one-dimensional stationary Schr\"odinger equation, which is provably non-degenerate\cite{Landau}. This property, however, clearly is very specific.  More generally, it may be necessary to consider two (or more) independent marginal modes, each having its own amplitude. In this scenario, the corresponding number of equations describing the slow-time evolution of each associated growth rate can be derived, and the algorithm described in this section adapted accordingly.

%%%% Insert B head here
%Subsection text here.

\section{Conclusion }

Multiple scales analysis is usually introduced to capture the dynamics of a single variable or field evolving over disparate (spatio-)temporal scales.  Canonical examples arise in the study of ordinary differential equations governing nonlinear oscillators (e.g., the van der Pol oscillator) and of partial differential equations governing nonlinear waves (e.g., the Korteveg-de-Vries equation).  Quasilinear (QL) partial differential equations constitute an important class of dynamical systems that requires a generalization of this approach to treat two (or more) tightly coupled fields evolving on different time scales.  

%It also reveals new physics that, fundamentally, requires at least two fields: wave/mean-flow coupling and the saturation of an instability through a feedback on the driving field.

Most (if not all) multi-scale QL systems can be categorized into one of the two model systems analyzed in this article. For systems involving wave/mean-flow interactions, and more generally for any QL system in which the fluctuations are slowly modulated waves, an amplitude equation can be derived by imposing an appropriate solvability condition on the dynamics of the correction fields.  As is well known\cite{Kruskal, Kuzman}, the amplitude equation reduces to the conservation of wave action if neither dissipation nor external forcing acts directly on the waves, in which case there exist complementary and, arguably, more elegant mathematical approaches -- including the generalized Lagrangian mean formalism \cite{Mcintyre,Buhler} and variational-based Hamiltonian methods\cite{Whitham, Salmon} --  to derive the appropriate conservation law(s) for the slow dynamics.  Nevertheless, these alternative methods cannot readily incorporate forcing and dissipation, which are naturally and straightforwardly included in the methodology presented here.  Moreover, the tight coupling between the wave and mean (e.g., celerity) fields renders our non-variational approach non-standard, as evidenced by the occurrence of \textit{functional derivatives} in the coefficients arising in the amplitude equation that account for changes in the leading-order wave eigenfunction resulting from the slow evolution of the leading-order mean field; nevertheless, these functional derivatives can be explicitly evaluated, so that costly sensitivity analyses are not required.

For many QL systems, particularly those arising in the modeling of spatially anisotropic turbulent shear flows (e.g., in wall-bounded engineering flows and in geo- and astrophysical turbulence\cite{Bouchet2013, Tobias2013, Chini_strat_turb, Keith, Squire}), the fluctuations can exhibit various forms of dynamical instability.  Generically, the QL reduction for such systems is only valid in the limit in which there is temporal scale separation and, in this limit ($\epsilon\to 0$ in our notation), the fluctuations can grow (or decay) exponentially on the fast time scale.  Importantly, for these systems, any attempt to derive an amplitude equation for the slow evolution of the fluctuations by employing the ``usual'' multiple scales approach (i.e., seeking a solvability condition by examining the equations for the correction fields) \emph{fails}, instead merely providing closures needed for evaluation of the corrections fields, should they be desired.  A primary contribution of the present article is the introduction of a new multiple scales formalism that is applicable to QL systems that exhibit dynamically-unstable fluctuation fields.  The essential point is that the slowly-evolving amplitude of the fluctuations must be instantaneously prescribed -- i.e., slaved to the slow mean field(s) -- to prevent positive growth rates from being realized once a state of marginal stability is attained.  In ongoing work, we are applying this new formalism to strongly stratified turbulent shear flows.

We conclude by emphasizing that multiple scales analysis can yield a sizable improvement in computational efficiency, as the time step employed by the numerical time-integrator can be increased from a fraction of a fast-time unit to a fraction of a slow-time unit.  Clearly, this reduction in computational expense is crucial for the study of several realistic physical systems exhibiting strong scale separation.  (In the quasi-biennial oscillation, for instance, the slow and fast time scales are separated by five orders of magnitude!)  Even for systems in which the scale separation is less extreme, QL models have proven useful in many applications\cite{Tobias2013, Squire, Thomas, Bretheim,  Tobias2017, Pausch}.  We emphasize that when dynamical instabilities are possible these QL systems \emph{must} self-tune toward a state of marginal stability, at least in a statistical (i.e., time-averaged) sense.  Thus, the analysis and algorithm developed in Sec.~\ref{sys2} should prove valuable even for modest values of $\epsilon$.\\

\enlargethispage{20pt}

%\ethics{Insert ethics text here.}

\dataccess{All data included in this article can be generated from the Python codes provided as supplementary material.}

\aucontribute{GM and GPC conceived  the  mathematical models, interpreted  the  results,  and  wrote  the
paper.  GM performed the numerical simulations.  Both  authors  gave  final  approval  for
publication.}

\competing{We declare we have no competing interests.}

\funding{This research was supported in part by the National Science Foundation under Grant No. NSF PHY-1748958.}

\ack{The authors are grateful to S. Tobias, C. Doering, and K. Julien for fruitful discussions, and acknowledge the hospitality of the Kavli Institute for Theoretical Physics at the University of California, Santa Barbara, where much of this research was completed.}

%\disclaimer{Insert disclaimer text here.}

%%%%%%%%%% Insert bibliography here %%%%%%%%%%%%%%

\vskip2pc

%\noindent {\bf Please follow the coding for references as shown below.}

%\noindent If maintaining .bib file for references, then please use "RS.bst" to generate the references.

%\noindent Example:

%\verb+\bibliographystyle{RS}+ %%%% .BST file

%\verb+\bibliography{sample}+ %%%%% .Bib file

\end{document}